\documentclass[11pt]{article}
\usepackage[textwidth=15.2cm,textheight=22cm]{geometry}
\usepackage{amsmath,amssymb}
\usepackage{latexsym}
\usepackage{multicol}
\usepackage{graphicx}
\usepackage{bm}
\allowdisplaybreaks[1]

\newcommand{\be}{\begin{equation}}
\newcommand{\ee}{\end{equation}}
\newcommand{\ba}{\begin{eqnarray}}
\newcommand{\ea}{\end{eqnarray}}
\newcommand{\bdm}{\begin{displaymath}}
\newcommand{\edm}{\end{displaymath}}

\newcommand\fr[1]{\frac{1}{#1}}

\def\ba{\bar A}

\def\beq{\begin{equation}}
\def\eeq{\end{equation}}
\newcommand{\half}{\frac{1}{2}}
\newcommand{\nn}{\nonumber}

\newcommand{\ndt}{\noindent}

\def\bea{\begin{eqnarray}}
\def\eea{\end{eqnarray}}
\def\beas{\begin{eqnarray*}}
\def\eeas{\end{eqnarray*}}
\def\sla{\raise.15ex\hbox{$/$}\kern-.57em}

\def\parp{\partial^+}

\def\spa#1.#2{\left\langle#1\,#2\right\rangle}
\def\spb#1.#2{\left[#1\,#2\right]}

\begin{document}

\begin{titlepage}
\begin{flushright}    
{\small $\,$}
\end{flushright}
\vskip 1cm
\centerline{\Large{\bf {Deriving spin-$1$ quartic interaction vertices}}}
\vskip 0.5cm
\centerline{\Large{\bf{from closure of the Poincar\'e algebra}}}
\vskip 1.5cm
\centerline{Sudarshan Ananth, Aditya Kar, Sucheta Majumdar and Nabha Shah}
\vskip .5cm
\centerline{\it {Indian Institute of Science Education and Research}}
\centerline{\it {Pune 411008, India}}
\vskip 1.5cm
\centerline{\bf {Abstract}}
\vskip .5cm
We derive the quartic interaction vertex of pure Yang-Mills theory by demanding closure of the light-cone Poincar\'e algebra in four-dimensional Minkowski spacetime. This calculation explicitly shows why structure constants must satisfy the Jacobi identity. We prove that there is no correction to the spin generator, for spin one, at this order. We comment briefly on higher spin fields in this context.
\vfill
\end{titlepage}

\section{Introduction}
\ndt In light-cone gauge, neither locality nor Poincar\'e invariance is manifest. Poincar\'e invariance thus needs to be explicitly verified  in light-cone field theories. As we will stress in this paper, this requirement of closure of the Poincar\'e algebra, apart from being an important check, may be viewed as a first-principles approach to deriving Lagrangians for interacting field theories in light-cone gauge.
\vskip 0.3cm
\ndt
The procedure involves starting with an ansatz for the interaction vertices and allowing the Poincar\'e  algebra to completely fix the ansatz. This algebraic approach has many advantages: at first order in the coupling constant, for theories involving fields of odd-spin the algebra requires the introduction of an antisymmetric constant~\cite{BBB}. In this paper, we explicitly show that the requirement of Poincar\'e invariance, at second order in the coupling constant, forces these constants to satisfy the Jacobi identity, signaling the emergence of a gauge group\footnote{In~\cite{OP}, similar results for the structure constants were derived using symmetry arguments in a covariant approach, starting with a general local relativistically-invariant Lagrangian.}.
\vskip 0.3cm
\ndt If we were to adapt this approach, for spins greater than $2$, to an AdS$_4$ background, would the closure of the SO($3,2$) algebra force the introduction of fields with spin larger than the one being considered? In principle, this could yield the elusive action principle corresponding to the Vasiliev equations~\cite{MV} (which would have to be established order by order in comparision with the Vasiliev model). It remains unclear whether the light-cone formalism presented here circumvents the many covariant no-go results~\cite{ngr}, pertaining to higher spin fields in flat spacetime, which assume both locality and Poincar\'e invariance to be manifest.
\vskip 0.3cm
\ndt In this paper, we extend the symmetry-based approach to deriving interaction vertices, introduced in~\cite{BBB}, to second order in the coupling constant for the specific case of spin one. We comment on similar expressions in the spin two and higher spin cases.
\vskip 0.5cm

\section{At order $\alpha$ : old results}
\vskip 0.3cm

Our metric signature is $(-,+,+,+)$ and we use light-cone coordinates
\begin{eqnarray}
x^{\pm}=\frac{x^{0}\pm x^{3}}{\sqrt{2}} \;,\qquad
x = \frac{x^{1}+ix^{2}}{\sqrt{2}} \;,\qquad\bar{x}= \frac{x^{1}-ix^{2}}{\sqrt{2}}\ ,
\end{eqnarray}
the accompanying derivatives being $\partial^{\mp}\,,\,\,\bar{\partial}$ and $\partial$.  Light-cone time is $x^+$ implying that $\partial^-$, the conjugate momentum, is the light-cone Hamiltonian.  Massless fields, in light-cone gauge, have two physical degrees of freedom $\phi$ and $\bar{\phi}$ corresponding to the $+$ and $-$ helicity states respectively. The generators of the Poincar\'{e} algebra are the momenta
\bea
P^-=-i\frac{\partial \bar\partial}{\parp}=-P_+ \qquad P^+=-i\partial^+=-P_- \qquad P=-i\partial \qquad \bar P=-i\bar\partial\ .
\eea
The $\frac{1}{\parp}$, an artifact of this gauge choice, is an integral operator acting on a field~\cite{SM}
\bea
\frac{1}{\parp}\phi (x^-)=\int\,dy^-\;\epsilon^{} (y^--x^-)\,\phi(y^-)\ ,
\eea
where $\epsilon$ is the step function. The rotation generators are
\begin{eqnarray}
J = (x \bar\partial -\bar x \partial - \lambda) \;,\qquad  J^+ = i(x\,\parp-x^+\partial)\ , \nn\\
J^{+-}=i(x^-\parp-x^+\frac{\partial \bar\partial}{\parp}) \;,\qquad J^{-}=i(x\frac{\partial \bar\partial}{\parp}-x^{-}\partial-\lambda \frac{\partial}{\parp}) \ ,
\end{eqnarray} 
and their complex conjugates. $\lambda$ is the spin of the field the generators act on. We work on the surface $x^+=0\,$ to simplify our calculations. The kinematical Poincar\'e generators do not involve time derivatives and include 
\bea
P^+\;,\;\;P\;,\;\;\bar P\;,\;\; J \;,\;\;J^+\;,\;\;{\bar J}^+ \;\;{\mbox {and}}\;\; J^{+-}\ .
\eea
The dynamical generators are 
\bea
P^-\;,\;\;J^-\;,\;\;{\bar J}^-\ ,
\eea
and pick up corrections when interactions are switched on. We introduce the Hamiltonian variation
\begin{eqnarray}
\label{hamvar}
\delta_{\mathcal H}\phi\equiv\partial^-\phi=\lbrace \phi,H\rbrace=\frac{\partial\bar\partial}{\parp}\,\phi\ ,
\end{eqnarray}
where the last equality only holds for the free theory. In the interacting theory, the $\delta_{\mathcal H}$ operator picks up corrections, order by order, in the coupling constant. 
\vskip 0.3cm
\ndt In the appendix, we provide a list of all the non-vanishing commutators satisfied by the light-cone Poincar\'e generators. The basic idea in~\cite{BBB}, that we follow in this paper, is to start with an ansatz for the operator $\delta_H\,\phi$, work through the entire list of Poincar\'e commutators to refine the ansatz and thus determine the Hamiltonian. For related discussions, see~\cite{poly2, poly1}.

\vskip 0.3cm
\subsection{Deriving cubic interaction vertices}
\vskip 0.3cm

\ndt We briefly review the derivation of cubic interaction vertices for integer $\lambda$. At this order the structure of the Hamiltonian is
\bea
\mathcal H \;\sim\; \alpha\;\bar\phi\,\phi\,\phi\;+\;\alpha\;\phi\,\bar\phi\,\bar\phi\ .
\eea
From (\ref {hamvar}), we see there are two contributions to $\delta_{\mathcal H}\,\phi$ at this order. The first involves two $\phi$ fields while the second involves a $\phi$ and a $\bar\phi$. We start with the following ansatz for the first variety 
\bea
\delta_{\mathcal H}^\alpha \phi=\alpha\,K\,\,{\parp}^\mu\,\biggl [ \,{\bar\partial}^B \,\partial^C\, {\parp}^\rho \, \phi\,\;{\bar\partial}^D\,\partial^E\,{\parp}^\sigma \phi \biggr ]\ ,
\eea
where $K$ is a constant and $\mu$, $\rho$, $\sigma$ $B,C,D,E$ are integers to be fixed by the algebra\footnote{The ansatz involving one $\phi$ and one $\bar\phi$ works very similarly.}. The commutator of this ansatz with $\delta_{J^{+-}}$ yields~\cite{BBB}
\bea
\label{delpc}
\mu+\rho+\sigma=-1\ .
\eea
We commute the ansatz with $\delta_J$ to obtain
\bea
B+D-C-E=\lambda\ .
\eea
Using (\ref {delpc}) and dimensional analysis we then find that
\bea
B+D=\lambda \quad ; \qquad C=E=0\ .
\eea
The other commutation relations determine the values of $\mu$, $\rho$ and $\sigma$. We thus obtain~\cite{BBB}
\begin{eqnarray}
\label{sc-1}
\delta_{\mathcal{H}}^{\alpha}\phi = \alpha \sum^{\lambda}_{n=0} (-1)^{n}{\lambda \choose n}{(\partial^+)}^{(\lambda -1)}\left[\frac{\bar{\partial}^{(\lambda -n)}}{\partial^{+(\lambda -n)}}\phi \frac{\bar{\partial}^{n}}{\partial^{+n}}\phi\right]\ ,
\end{eqnarray}
for even $\lambda$. 
\vskip 0.3cm
\subsection*{Appearance of the ``structure constant"}
\vskip 0.3cm
\ndt Interestingly, a non-trivial solution for odd-helicity fields is only possible through the introduction of an antisymmetric three-index object $f^{abc}$,
\begin{equation}
\label{sc}
\delta_{\mathcal{H}}^{\alpha}\phi^{a} = \alpha f^{abc}\sum^{\lambda}_{n=0} (-1)^{n}{\lambda \choose n}{(\partial^+)}^{(\lambda -1)}\left[\frac{\bar{\partial}^{(\lambda -n)}}{\partial^{+(\lambda -n)}}\phi^{b} \frac{\bar{\partial}^{n}}{\partial^{+n}}\phi^{c}\right]\ .
\end{equation}
\ndt The same procedure determines $\delta_H$ corresponding to the $\alpha\, \bar \phi \phi$ structure. The Hamiltonian, to this order, follows from (\ref {sc-1}) and (\ref {sc}). The corresponding actions read~\cite{BBB}
\be
\label{even}
S=\int d^{4}x \left( \half \bar{\phi}\Box\phi+\alpha \sum^{\lambda}_{n=0} (-1)^{n}{\lambda \choose n}\bar{\phi}{(\partial^+)}^{\lambda}\left[\frac{\bar{\partial}^{(\lambda -n)}}{\partial^{+(\lambda -n)}}\phi \frac{\bar{\partial}^{n}}{\partial^{+n}}\phi\right]+c.c. \right),
\ee
for even $\lambda$ and
\be
\label{odd}
S=\int d^{4}x  \left( \half \bar{\phi}^{a}\Box\phi^{a}+\alpha f^{abc}\sum^{\lambda}_{n=0} (-1)^{n}{\lambda \choose n}\bar{\phi}^{a}{(\partial^+)}^{\lambda}\left[\frac{\bar{\partial}^{(\lambda -n)}}{\partial^{+(\lambda -n)}}\phi^{b} \frac{\bar{\partial}^{n}}{\partial^{+n}}\phi^{c}\right]+c.c. \right) ,
\ee
for odd $\lambda$.
\vskip 0.3cm
\ndt It is interesting to note~\cite{AKHB} that, at the cubic level, the non-linear dynamical part of the algebra does not restrict the cubic interactions beyond what is required by the kinematical portion of the algebra (see also~\cite{BC}).
\vskip 0.5cm

\section{At order $\alpha^2$ : new results}

\ndt In this section, we extend this formalism to second order in the coupling constant for the specific case of $\lambda=1$. The two fields in Yang-Mills theory, $A$ and $\bar A$, have helicity $+1$ and $-1$ respectively. We also identify $2\,\alpha$ with the dimensionless Yang-Mills coupling constant $g$ . We will use the following result from the previous section,
\bea
\delta_H^g A^a=+g\,f^{abc}\,\biggl \{ -A^c\,\frac{\bar\partial}{\parp}\,A^b + \fr{{\parp}^2}({\parp}^2 A^b\,\frac{\partial}{\parp}{\bar A}^c)-\fr{{\parp}^2}(\partial \parp A^b\,{\bar A}^c)\ \biggr \}.
\eea
\ndt We will also need the corrections to the spin generator at this order~\cite{BBB}
\bea
\delta_{{\bar S}^-}^g A^a=-g\,f^{abc}\,\fr{{\parp}^2} \biggl ( \fr{\parp}\,{\bar A}^c\,{\parp}^2A^b+3\,{\bar A}^c \parp A^b \biggr )\ ,
\eea

\bea
\delta_{S^-}^g A^a=+g\,f^{abc}\,\fr{\parp}\,A^b \, A^c\ .
\eea

\vskip 0.3cm
\ndt The key commutator involving dynamical generators is
\bea
\label{main1}
[\delta_{J^-}\,,\,{\delta_H}]\,A^a=0\ .
\eea
\vskip 0.1cm
\ndt In the following computation, we present only terms of the form $AA\bar A$ (terms of the form $AAA$ vanish independently). We note that the other dynamical commutator between $\delta_{J^-}$ and $\delta_{{\bar J}^-}$ does not yield additional information because $\delta_{{\bar J}^-}A^a$ at order $g^2$ is proportional to ${\delta_H}A^a$ (see appendix B in \cite {poly2} for related discussions). We begin by computing contributions to (\ref {main1}) at order $g^2$, from $[\delta_{J^-}^g\,,\,{\delta_H}^g]\,A^a$. This calculation involves two pieces, orbital and spin,
\bea
\label{gg1}
[\delta_{L^-}^g\,,\,{\delta_H}^g]\,A^a=[x\,{\delta_H}^g\,,\,{\delta_H}^g]\,A^a=-g\,f^{abc}\,A^c\fr{\parp}({\delta_H}^g\,A^b)\ ,
\eea
and
\bea
\label{gg2}
[\delta_{S^-}^g\, ,\, \delta_H^g ]\, A^a=&&\!\!\!\!\!\!\!\!\!\!+\,g^2\,f^{abc}\, \biggl \{ f^{bde}\fr{{\parp}^2}{\biggl (}{\parp}^2(\fr{\parp}A^d\,A^e)\frac{\partial}{\parp}{\bar A}^c \biggr )\!\!-\!\!f^{bde}\fr{{\parp}^2}{\biggl (}\partial\parp (\fr{\parp}A^d\,A^e){\bar A}^c \biggr ) \nn \\
&&-f^{cde}\,\fr{{\parp}^2}{\biggl (}{\parp}^2\,A^b\,\frac{\partial}{{\parp}^3}(\fr{\parp}A^e\,{\parp}^2{\bar A}^d+3A^e\parp {\bar A}^d) \biggr )\!\! \nn \\
&&+f^{cde}\,\fr{{\parp}^2}{\biggl (}\partial \parp\,A^b\,\fr{{\parp}^2}(\fr{\parp}A^e\,{\parp}^2{\bar A}^d+3A^e\parp {\bar A}^d) \biggr )\,\biggr \} \nn \\
&&-g\,f^{abc}\,\delta_H^g A^c\,\fr{\parp}A^b-g\,f^{abc}\,A^c\,\fr{\parp}(\delta_H^g\,A^b)\ .
\eea
\ndt The other contribution to (\ref {main1}) is from commutators that involve one generator at order $g^0$ and one at order $g^2$. Before we evaluate these, we need an ansatz for $\delta_H^{g^2}$. We begin with a very general structure that is the {\it {sum}} of terms of the form\footnote{One may write down other combinations by moving the derivatives around but these structures can be generated starting from the form in (\ref {ansz}).}
\bea
\label{ansz}
\delta_H^{g^2} A^a=+g^2\,K\,f^{abc}\,f^{cde}\,{\parp}^\mu\,\biggl [ \,{\bar\partial}^B\,\partial^C\, {\parp}^\rho A^b\,\;{\parp}^\sigma \biggl (\,{\bar\partial}^D\,\partial^E\,{\parp}^\eta{A^d}\;\,{\bar\partial}^F\,\partial^G\,{\parp}^\delta\,{\bar A}^e\,\biggr )\; \biggr ]\ ,
\eea
where $K$ is a constant and $\mu$, $\rho$, $\sigma$, $\eta$, $\delta$, $B,C,D,E,F,G$ are integers to be determined by the algebra. We commute this with $\delta_J$ to find the following conditions.
\bea
B+D+F=C+E+G=\lambda-1\ .
\eea
Thus no transverse derivatives are permitted when $\lambda=1$, simplifying our ansatz to
\bea
\label{simans}
\delta_H^{g^2} A^a=+g^2\,K\,f^{abc}\,f^{cde}\,{\parp}^\mu\,\biggl [ \,{\parp}^\rho \, A^b\,\;{\parp}^\sigma \biggl (\,{\parp}^\eta{A^d}\;\,{\parp}^\delta\,{\bar A}^e\,\biggr ) \; \biggr ]\ .
\eea
The commutator with $\delta_{J^{+-}}$ yields
\bea
\label{delpcond}
\mu+\rho+\sigma+\eta+\delta=-1\ .
\eea
The final piece of the computation involves
\bea
\label{comp7}
[\delta_{L^-}^{g^2},\delta_H^0]\,A^a+[\delta_{J^-}^{0},\delta_H^{g^2}]\,A^a\ ,
\eea
where we have ignored the spin generator in the first commutator (since it is zero as explained in the next subsection). We find that the following solution
\bea 
\label{values}
(\mu=-1\,; \rho=+1\,; \sigma=-2\,; \eta=0\,; \delta=+1)\!+\!(\mu=0\,; \rho=0\,; \sigma=-2\,; \eta=+1\,; \delta=0)\ ,
\eea
satisfies (\ref {main1}) and present below the explicit computation of (\ref {comp7}) for these values (the more general case is far more lengthy but not necessary for the points we wish to make). Any other set of consistent values for these constants is completely equivalent to those above (essentially corresponding to trivial re-writings of the result in (\ref {res})).

\bea
\label{gsq}
&&\!\!\!\!\!\!f^{abc}f^{cde}[-\frac{1}{\partial^{+2}}({ \partial^{+}\partial  A^b \frac{1}{\partial^{+2}}}({\bar{A}}^{e}\partial^{+} A^{d}))+\frac{1}{\partial^{+2}}({ \partial^{+}\partial  A^b \frac{1}{\partial^{+2}}}({\partial^{+}\bar{A}}^{e} A^{d}))+\frac{1}{\partial^{+2}}( \frac{\partial}{\partial^{+}}  A^b\bar{A}^{e}\partial^{+} A^{d}) \nn \\
&&\!\!\!\!\!\!+2 \frac{1}{\partial^{+2}}(\partial^{+}A^{b} \frac{1}{\partial^{+2}}(\bar{A}^{e} \partial\partial^{+} A^{d}))- \frac{1}{\partial^{+2}}(\partial^{+}A^{b}\frac{1}{\partial^{+}}(\partial^{+}\bar{A}^{e} \frac{\partial}{\partial^{+}}A^{d})) - 2 \frac{1}{\partial^{+2}}(\partial^{+}A^{b} \frac{1}{\partial^{+}}(\frac{\partial}{\partial^{+}} \bar{A}^{e} \partial^{+}A^{d})) \nn \\
&&\!\!\!\!\!\!+\frac{1}{\partial^{+2}}(\partial^{+}A^{b} \frac{1}{\partial^{+2}}(\partial^{+}\partial \bar{A}^{e} A^{d}))-2\frac{1}{\partial^{+2}}(\partial^{+2}A^{b}\frac{1}{\partial^{+3}}(\partial^{+} \bar{A}^e \partial A^{d}))+4\frac{1}{\partial^{+2}}(\partial^{+2}A^{b}\frac{1}{\partial^{+3}}(\partial \bar{A}^e \partial^{+} A^{d})) \nn \\
&&\!\!\!\!\!\!+2\frac{1}{\partial^{+2}}(\partial^{+2}A^{b}\frac{1}{\partial^{+3}}( \bar{A}^e \partial \partial^{+} A^{d}))-\frac{1}{\partial^{+2}}(\partial^{+2}A^{b}\frac{1}{\partial^{+2}}(\partial^{+} \bar{A}^e \frac{\partial}{\partial^{+}} A^{d}))-\frac{1}{\partial^{+2}}(\partial^{+2}A^{b}\frac{1}{\partial^{+2}}(\frac{\partial}{\partial^{+}} \bar{A}^e \partial^{+} A^{d})) \nn \\
&&\!\!\!\!\!\!-\frac{1}{\partial^{+2}} (A^{b}\frac{\partial}{\partial^{+}} \bar{A}^{e}\partial^{+} A^{d})- \frac{1}{\partial^{+2}}(\partial^{+}A^{b} \frac{1}{\partial^{+2}}(\partial^{+} \bar{A}^{e} \partial A^{d}))+4 \frac{1}{\partial^{+2}}(\partial^{+}A^{b} \frac{1}{\partial^{+2}}(\partial \bar{A}^{e} \partial^{+} A^{d}))]\ .
\eea
\vskip 0.3cm
\subsection*{Emergence of a gauge group}
\vskip 0.3cm
\ndt The crucial point here is that the two expressions in (\ref {gg1}) and (\ref {gg2}) cancel perfectly against (\ref {gsq}) {\it {if and only if}} we assume that the $f^{abc}$ introudced in (\ref {sc}) satisfy the Jacobi identity$\,$\footnote{The Jacobi identity is also necessary to prove that terms of the form $AAA$ vanish.},
\bea
f^{abc}\,f^{bde}+\,f^{abd}\,f^{bec}+f^{abe}\,f^{bcd}\,=\,0\ .
\eea
Thus, we find
\bea
\label{res}
\delta_H^{g^2} A^a=g^2\,f^{abc}\,f^{cde}\,\biggl [ \fr{\parp} \biggl ( \parp A^b \,\fr{{\parp}^2} (\parp {\bar A}^e A^d) \biggr ) - A^b\,\fr{{\parp}^2} ( {\bar A}^e \, \parp A^d ) \biggr ]\ .
\eea
This expression leads to the same quartic interaction vertex obtained by light-cone gauge-fixing the covariant Yang-Mills Lagrangian~\cite{BLN}.

\vskip 0.3cm
\subsection{The spin generator at order $g^2$}
\vskip 0.1cm
\ndt To show that $\delta_{S^-}^{g^2}\,A^a=0$, we start by examining the helicities and dimensions involved.
\begin{center}
\begin{tabular}{|c c c|} 
\hline
Quantity & Helicity & Dim $[L]$ \\ [0.5ex]
\hline 
$x$ & $+1$ & $+1$ \\ [1.2ex] 
$\bar x$ & $-1$ & $+1$ \\ [1ex]
$\partial$ & $+1$ & $-1$ \\ [1ex]
$\bar\partial$ & $-1$ & $-1$ \\ [1ex]
$A$ & $+1$ & $-1$ \\ [1ex]
$\bar A$ & $-1$ & $-1$ \\ [1ex]
$\parp$ & $\;\;\,0$ & $-1$ \\ [0.3ex] 
\hline
\end{tabular}
\end{center}
\vskip 0.2cm
At lowest order,
\bea
\delta_{S^-}^0 A^a=-\frac{\partial}{\parp}\,A^a\ ,
\eea
has helicity $+2$ and a length-dimension of $-1$. An ansatz at order $g^2$ will take the form
\bea
\label{guess1}
\delta_{S^-}^{g^2} A\,\sim\,g^2\,A\,A\,\bar A\,\partial \,\fr{{\parp}^3}\ ,
\eea
where the derivatives at the end of the expression may be sprinkled on various fields. However, the commutator with $\delta_{J^{+-}}$ works only if the number of $\parp$'s in the denominator is one greater than that in the numerator (see for example (\ref {delpcond})) ruling out this ansatz. No combination of ingredients from the table above, with three fields, has the correct values of helicity, dimension and kinematical commutators. The same argument rules out the possibility of a non-zero $\delta^{g^2}_{{\bar S}^-}A^a$. This is not surprising since the Poincar\'e commutator $[\bar P, J^-]=-i P^-$ tells us that the spin generator has one less transverse derivative than the Hamiltonian (and the spin one Hamiltonian at order $g^2$ has zero transverse derivatives).
\vskip 0.3cm
\ndt {\it {With this, the construction of the entire light-cone Poincar\'e algebra for Yang-Mills theory is complete.}}

\section{Comments}
\vskip 0.3cm

\ndt We briefly examine $\delta_H$ for the case of $\lambda=2$. At lowest order
\bea
\delta_H^0 h=\frac{{\partial}{\bar \partial}}{\parp}\,h\ ,
\eea
this has a length-dimension of $-2$ and a helicity of $+2$. At order $\alpha^2$ we expect the form
\bea
\label{form}
\delta_H^{\alpha^2} h \,\sim\,\alpha^2\,h h \bar h\ ,
\eea
where $\alpha$ has the dimensions of length, for a spin two field. The structure in (\ref {form}) has the correct helicity but the wrong dimension. Using constraints like (\ref {delpcond}) and dimensional analysis, we conclude that
\bea
\label{formf}
\delta_H^{\alpha^2} h \,\sim\,\alpha^2\,h h  \bar h\;\,(\partial\bar\partial)\,\,\fr{\parp}\ .
\eea
\ndt Note that we can introduce equal numbers of $\parp$ and $\fr{\parp}$ through-out the expression resulting in a sum of terms. This matches the structures that appear from gauge-fixing the gravity Lagrangian~\cite{ABHS}.
\vskip 0.3cm

\vskip 0.3cm
\ndt At next order, we expect
\bea
\delta_H^{\alpha^3} h \,\sim\,\alpha^3\,h h \bar h \bar h\;\partial^2\,\fr{\parp}+\alpha^3\,h h h \bar h\;{\bar \partial}^2\,\fr{\parp}\ ,
\eea
which again precisely matches the structure written down in~\cite{SA2}. The sheer volume of terms involved in these expressions at orders $\alpha^2$ and $\alpha^3$ make algebraic computations tedious to work out for spins $\geq 2$. 
\vskip 0.1cm
\ndt For higher spin fields
\bea
\delta_H^{\alpha^2} \phi \,\sim\,\alpha^2\,\phi \phi  \bar \phi\;{(\partial\bar\partial)}^{\lambda-1}\,\fr{\parp}\ ,
\eea
\ndt where $\alpha$ now has length-dimension of $\lambda-1$. The dynamical commutators, in this case, would play a key role in determining whether consistent quartic vertices are allowed in flat spacetime. The corresponding corrections to the spin generator at this order read
\bea
\delta_{S^-}^{\alpha^2} \phi\,\sim\,\alpha^2 \,\phi\,\phi\, \bar \phi\, \partial \,{(\partial\bar\partial)}^{\lambda-2}\,\fr{\parp}\ .
\eea
\vskip 0.3cm

\begin{center}
* ~ * ~ *
\end{center}

\vskip 0.3cm

\ndt The preceeding section raises the question of whether higher spin fields can have consistent interacting vertices in flat spacetime, within this formalism.  For related discussions, see~\cite{RRM}. An obvious next step is to modify this formalism and apply it on an AdS background~\cite{RRM2}. Just as the algebra, in flat spacetime, led us to structure constants and the Jacobi identity, it is plausible that the same approach on an AdS background~\cite{AAM} applied to a spin $>2$ field will lead us to re-discover the higher spin tower.

\vskip 0.5cm
\ndt {\it {Acknowledgments}}
\vskip 0.3cm

\ndt We thank Anders Bengtsson for helpful discussions. We thank the anonymous referee for valuable suggestions. AK and NS acknowledge support from DST Inspire fellowships. The work of SM is supported by a CSIR NET fellowship. The work of SA is partially supported by a DST-SERB grant (EMR/2014/000687). 

\newpage
\appendix
\section{Light-cone Poincar\'e algebra}

\ndt We define 

\be
J^+\ =\ \frac{J^{+1}+iJ^{+2}}{\sqrt{2}}\ ,\quad \bar J^{+}\ =\ \frac{J^{+1}-iJ^{+2}}{\sqrt 2}\ ,\quad J\ =\ J^{12}\ .
\ee
\vskip 0.2cm
\ndt
The {\it {non-vanishing}} commutators of the Poincar\'e algebra are

\bea
&[P^-, J^{+-}]\ =\ -i P^- \ , \quad  &[P^-, J^+] = -i P\ , \quad [P^-, \bar J^+]\ =\ -i \bar P \nn \\ \nn\\
&[P^+, J^{+-}]\ = \ iP^+\ ,\quad  &[P^+, J^-]\ =\ -iP\ , \quad [P^+, \bar J^-]\ =\ -i \bar P \nn \\ \nn \\
&[P, \bar J^-]\ =\ -i P^-\ ,  \quad &[P, \bar J^+] \ =\ -iP^+\ , \quad [P, J]\ =\ P \nn \\ \nn \\
&[\bar P, J^-]\ = \ -i P^-\ , \quad &[\bar P, J^+]\ =\ -iP^+\ , \quad [\bar P, J]\ =\ -\bar P \nn \\ \nn \\
&[J^-, J^{+-}]\ =\ -i J^- \ , \quad &[J^-, \bar J^+]\ =\ iJ^{+-} +  J \ , \quad [J^-, J]\ =\ J^- \nn \\ \nn \\
&[\bar J^-, J^{+-}]\ = \ -i \bar J^- \ , \quad &[\bar J^-, J^+]\ =\ iJ^{+-} - J \ , \quad [\bar J^-, J]\ =\ -\bar J^- \nn \\ \nn \\
&[J^{+-}, J^+]\ =\ -i J^+ \ ,  \quad &[J^{+-}, \bar J^+]\ = \ -i \bar J^+ \ , \nn \\ \nn \\
&[J^+, J]\ =\ J^+\ , \quad &[\bar J^+, J]\ =\ -\bar J^+\ .
\eea


\newpage

\begin{thebibliography}{Ref}
\bibitem{BBB}{A. K. H. Bengtsson, I. Bengtsson and L. Brink, {\it Nucl. Phys.} {\bf B 227} (1983) 31. \\ 
A. K. H. Bengtsson, I. Bengtsson and N. Linden, {\it Class. Quant. Grav.} {\bf 4} (1987) 1333.}
\bibitem{OP}{V. I. Ogievetskij and I. V. Polubarinov, {\it Ann. Phys.} {\bf 25} (1963) 358.}
\bibitem{MV}{E. S. Fradkin and M. A. Vasiliev, {\it Nucl. Phys.} {\bf B 291} (1987), 141. \\
M. A. Vasiliev, {\it Phys. Lett.} {\bf B 243}, 378 (1990). \\
M. A. Vasiliev, {\it Phys. Lett.} {\bf B 285}, 225 (1992).}
\bibitem{ngr}{S. Weinberg and E. Witten, {\it Phys. Lett.} {\bf B 96} (1980) 59.\\
G. Velo and D. Zwanziger, {\it Phys. Rev.} {\bf 188} (1969) 2218. \\
S. R. Coleman and J. Mandula, {\it Phys. Rev.} {\bf 159} (1967) 1251. \\
A. K.H. Bengtsson and I. Bengtsson, {\it Class. Quant. Grav.} {\bf 3} (1986) 927.\\
A. K. H. Bengtsson, C15-11-04.1, 353, (2016) arXiv:1605.02608. \\
A. K. H. Bengtsson, {\it JHEP} {\bf 1612} (2016) 134, arXiv:1607.06659.}
\bibitem{SM}{S. Mandelstam, {\it Nucl. Phys.} {\bf B 213} (1983) 149.
J. D. Bjorken, J. B. Kogut and D. E. Soper, {\it Phys. Rev.} {\bf D 3} (1971) 1382.}
\bibitem{poly2}{D. Ponomarev and E. D. Skvortsov, {\it J. Phys.} {\bf A 50} (2017) 095401, arXiv:1609.04655.}
\bibitem{poly1}{D. Ponomarev, {\it JHEP} {\bf 1612} (2016) 117, arXiv:1611.00361.}
\bibitem{AKHB}{A. K. H. Bengtsson, arXiv:1205.6117 (2012).}
\bibitem{BC}{P. Benincasa and F. Cachazo, arXiv:0705.4305 (2007).}
\bibitem{BLN}{L. Brink, O. Lindgren and B. E.W. Nilsson, {\it Nucl. Phys.} {\bf B 212} (1983) 401.}
\bibitem{ABHS}{S. Ananth, L. Brink, R. Heise, H. G. Svendsen, {\it Nucl. Phys.} {\bf B 753} (2006) 195, arXiv:hep-th/0607019.}
\bibitem{SA2}{S. Ananth, {\it Phys. Lett.} {\bf B 664} (2008) 219, arXiv:0803.1494.}
\bibitem{RRM}{R. R. Metsaev, {\it Mod. Phys. Lett.} {\bf A 6} (1991) 359. \\
R. R. Metsaev, {\it Mod. Phys. Lett.} {\bf A 6} (1991) 2411.}
\bibitem{RRM2}{R. R. Metsaev, {\it Nucl. Phys.} {\bf B 563} (1999) 295. \\
R. R. Metsaev, {\it Int. J. Mod. Phys.} {\bf A 16S1C} (2001) 994-997.}
\bibitem{AAM}{Y.S. Akshay, S. Ananth and M. Mali, {\it Nucl. Phys.} {\bf B 884} (2014) 66, arXiv:1401.5933.}



 \end{thebibliography}
\end{document}